\documentclass[sigconf,screen,nonacm]{acmart}

\usepackage{listings}
\usepackage{graphicx}
\usepackage{svg}
\usepackage{algorithm}
\usepackage{algpseudocode}
\usepackage{multicol}
\usepackage{multirow}

\usepackage{caption}
\usepackage{ragged2e}
\usepackage{wrapfig}

\usepackage{tikz}
\usetikzlibrary{positioning, shapes.geometric}
\usepackage{tikz}
\usepackage{graphicx}
\usepackage{listings}
\usepackage{subcaption}

\usetikzlibrary{arrows.meta} 

\AtBeginDocument{%
  \providecommand\BibTeX{{%
    \normalfont B\kern-0.5em{\scshape i\kern-0.25em b}\kern-0.8em\TeX}}}

\setcopyright{acmcopyright}
\copyrightyear{2018}
\acmYear{2018}
\acmDOI{XXXXXXX.XXXXXXX}

\acmConference[Conference acronym 'XX]{Make sure to enter the correct
  conference title from your rights confirmation emai}{June 03--05,
  2018}{Woodstock, NY}
\acmPrice{15.00}
\acmISBN{978-1-4503-XXXX-X/18/06}

\definecolor{patriarch}{rgb}{0.5, 0.0, 0.5}
\definecolor{darkolivegreen}{rgb}{0.33, 0.42, 0.18}
\lstdefinestyle{cpp}{
language=C++,
basicstyle=\footnotesize,
keywordstyle=\color{patriarch},
commentstyle=\color{darkolivegreen},
stringstyle=\color{red},
showstringspaces=false,
breaklines=true,
frame=none,
numbers=left,
xleftmargin=0.4em,
numbersep=1pt,
numberstyle=\scriptsize\color{gray},
escapechar=|
}

\begin{document}

\title{A Novel Compiler Transformation for Fast Sparse Matrix Multiplication in GPUs}
\author{Hossein Albakri}
\orcid{1234-5678-9012}
\affiliation{
  \institution{McMaster University} 
  \city{Hamilton}
  \state{Ontario}
  \country{Canada}                 
}
\email{albakrih@mcmaster.ca}

\author{Kazem Cheshmi}                               
\orcid{0000-0002-2968-5176}             
\affiliation{
  \institution{McMaster University} 
  \city{Hamilton}
  \state{Ontario}
  \country{Canada}                 
}
\email{cheshmi@mcmaster.ca}


\begin{abstract}
Sparse data structures are commonly used in neural networks to reduce the memory footprint. These data structures are compact but cause irregularities such as random memory accesses, which prevent efficient use of the memory hierarchy. GPUs are a common platform for machine learning practitioners, but running compact data structures on these devices often leads to slow-downs due to inefficient use of computing and memory resources. This paper proposes a new compiler transformation, enumerate-and-sparse-coarsen, that accelerates sparse matrix-matrix multiplication (SPMM) on GPU devices. The transformation increases data reuse in registers and caches while creating more balanced workloads for GPU computing resources. The transformation is tested on sparse neural networks in convolutional and transformer models. 
On an A100 GPU and across a columns of matrix B (bCols) in $ A \times B = C$ from range of 32 to 128, the transformation yields a geometric mean speedup of 1.84$\times$ to 2.27$\times$ compared to cuBLAS and cuSPARSE baselines, respectively.
\end{abstract}





\received{20 February 2007}
\received[revised]{12 March 2009}
\received[accepted]{5 June 2009}

\maketitle

\section{Introduction}


As neural networks continue to grow in size, sparse neural networks have emerged as a promising solution to reduce their computational and memory footprint.
Neural network layers are often represented as matrix multiplications of the form $C=A\times B$, where $A$ is the weight matrix, $B$ is the input data or the output of the previous layer, and $C$ is the output of the current layer.
Large Deep Neural Networks (DNNs) are often over-parameterized~\cite{hoefler2021sparsity}, meaning they contain more parameters than necessary. Research has shown that DNNs can maintain good accuracy even after removing a significant portion of their parameters~\cite{hoefler2021sparsity}. This process, known as pruning, can be achieved through various algorithms.
One pruning technique involves removing individual weights based on their magnitude, resulting in matrices with element-wise sparsity, denoted as unstructured sparsity. 
Efficiently performing matrix multiplication on sparse weight matrices remains a challenging problem.

Dense matrix multiplication (MatMul) is commonly used for the training and inference of sparse neural networks on GPUs.
To fully utilize the potential of GPUs, the dense MatMul is broken into smaller MatMuls, known as tiles, to fit into GPU fast memory. Tiles are then mapped to a group of threads, i.e., \textit{thread-blocks}. Each thread block will be scheduled to one Streaming Multiprocessor (SM) in the GPU. GPUs have several SMs, each composed of several computing units, a.k.a., CUDA cores. Each MatMul thread is eventually executed by fused-multiply-add (FMA) functional units.
Since memory accesses in dense MatMul are known at compile time, they can be efficiently mapped to GPU functional units.
Thus, sparse neural networks are often mapped to efficient dense MatMul implementations such as cuBLAS~\cite{naumov2010cusparse}. While efficient, GPU resources will be wasted due to operations on zero elements.



Compressed sparse formats are commonly used for sparse MatMul, such as in SPMM, to prevent operations on zero elements. These formats, such as compressed sparse row (CSR), only store nonzero elements and keep track of nonzero coordinates with additional index arrays. Due to compaction, accessing nonzero elements is done through index arrays, creating indirect accesses. The indirect accesses in sparse codes hide memory accesses at compile-time, creating challenges for load balancing and locality.
Prior SPMM optimization efforts have focused on achieving load balance across threads through balanced tiling techniques. Load balancing is often done through either 1D tiling~\cite{gale2020sparse}, commonly supported by sparse formats, or 2D tiling~\cite{xiaflash,lin2023ec}, where sparsity information is used. While load balancing for sparse codes on GPUs has been extensively studied, thread-level optimization has received less attention.

It is known that efficient thread-level optimizations, such as instruction-level parallelism and register reuse within a thread, are essential for GPU performance~\cite{volkov2010better}. Thread-level optimization and load balancing are often competing factors in sparse MatMul optimization. 
Using thread-level optimization, especially for memory-bound sparse operations, is crucial for hiding memory latency. One common technique to enhance thread-level optimization is thread coarsening, which can improve register reuse and instruction-level parallelism. Specifying the degree of coarsening is complex on GPUs, as they have a large number of registers and coarsening can impact occupancy.
Coarsening techniques have been successfully applied in various applications~\cite{magni2014automatic}. Finding an appropriate coarsening strategy for sparse matrix multiplication is particularly challenging due to limited knowledge of access patterns at compile time.
We propose enumerate-and-sparse-coarsen, a new compiler transformation to map iterations of SPMM to GPU threads. The transformation relies on an enumeration step to ensure load balance using sparsity information and to prevent thread divergence.
It also uses a new coarsening approach, sparse coarsen, that ensures thread-level optimization such as register reuse and efficient use of caches and memory. The transformation is implemented as a source-to-source code generator that generates CUDA code from Python. It achieves speedups of 1.47$\times$ and 1.7$\times$ over cuBLAS and cuSparse, respectively, for bCol size of 32, 64 for the Deep Learning Matrix Collection (DLMC) \cite{gupta2024splat}, all measured on an NVIDIA A100 GPU.
%

\begin{figure}
    \centering
    \begin{subfigure}[b]{0.5\textwidth} 
        \centering
        \includegraphics[width=\linewidth]{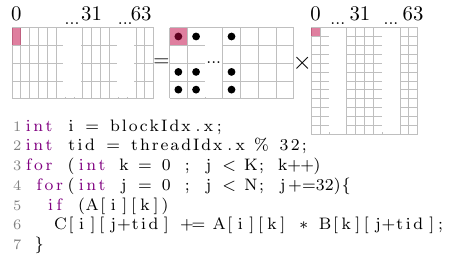}
        \caption{UFi=1, UFk=1, GridSize=M, ThreadBlockSize=32}
        \label{fig:motivc0}
    \end{subfigure}%
    \hfill 
    \begin{subfigure}[b]{0.5\textwidth} 
        \centering
        \includegraphics[width=\linewidth]{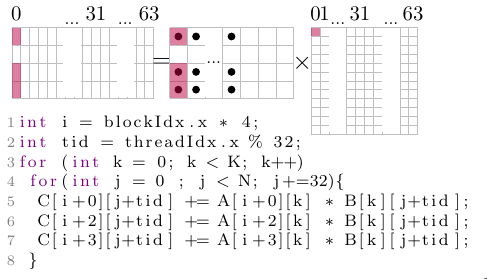}
        \caption{UFi=4, UFk=1, GridSize=M/4, ThreadBlockSize=32}
        \label{fig:motivc1}
    \end{subfigure}
    \hfill 
    \begin{subfigure}[b]{0.5\textwidth} 
        \centering
        \includegraphics[width=\linewidth]{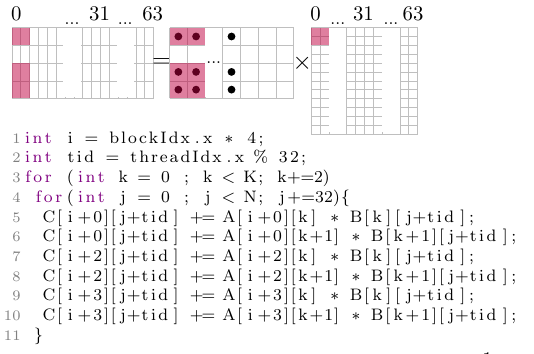}
        \caption{UFi=4, UFk=2, GridSize=M/4, ThreadBlockSize=32}
        \label{fig:motivc2}
    \end{subfigure}
    \caption{The effect of sparse-coarsening on a tile from a sparse matrix under different Unrolling Factors(UF). The sparse matrix is shown in a dense format and dark circles represent nonzero elements. Colored regions in matrices shows what operands are loaded for a thread in a thread block. All stores are atomic and not shown in the code.  }
    \label{fig:motiv}
\end{figure}

\begin{figure}[t]
    \centering
    \includegraphics[width=\linewidth]{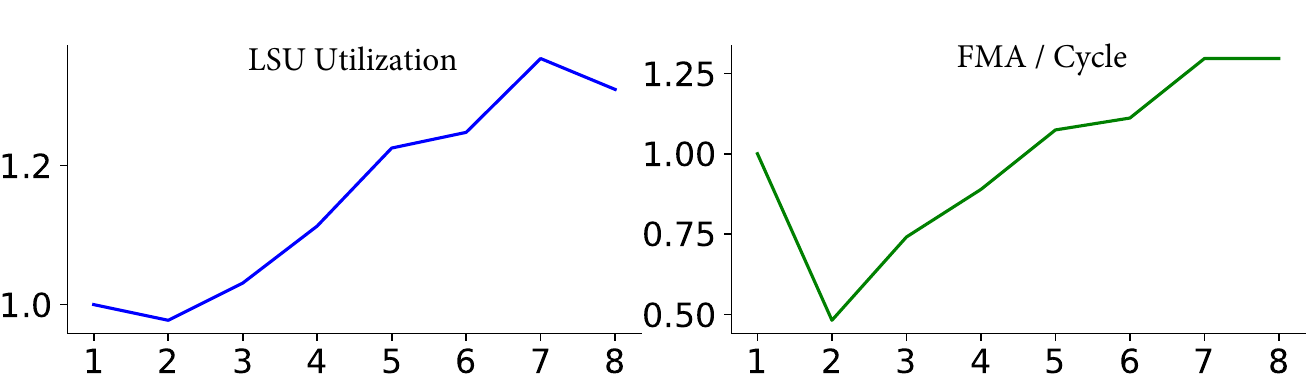}
    \caption{Normalized LSU utilization and FMA/cycle for different coarsening for a 1024$\times$256 matrix}
    \label{fig:motive_register}
\end{figure}

\section{Motivation} 

We use the SpMM shown in Figure~\ref{fig:motiv} to motivate thread coarsening and its optimization space in matrix multiplications. Figure~\ref{fig:motiv} presents three different thread mappings on a GPU, considering a sparse matrix $A$ of size $M \times K$ multiplied with a dense matrix $B$ of size $K \times N$. The result is stored in a dense matrix $C$ of size $M \times N$. 
The matrix is preprocessed to partition the tile into sub-tiles with identical sparsity patterns. For a tile size of 4, there are 15 possible sparsity patterns in the matrix $A$. 
For simplicity, only one sub-tile with three nonzeros is shown in Figure~\ref{fig:motiv}. 

Three different thread-level mappings are presented, each with varying levels of FMA utilization per thread. In each mapping, highlighted regions in the matrices represent non-zero locations that are accessed to load FMA operands in a thread within a thread block. All three codes are launched with a 1D grid and a thread block size of $32$. Highlighted nonzeros in $A$ in each figure correspond to an FMA statement in its listing.
We assume the underlying GPU first loads FMA operands from L1 to registers using a load/store unit (LSU). Then it executes them using an FMA unit. For operands already in the register, loading from the L1 cache is not needed, thus enabling register reuse and more efficient use of the LSU. Additionally, since operands are loaded into registers, more FMA instructions can be scheduled, potentially improving instruction-level parallelism. 


In Figure~\ref{fig:motivc0}, each thread block computes one row of $C$ as shown in line~6 in Figure~\ref{fig:motivc0}. 
It also demonstrates how each iteration of the \texttt{j} loop is mapped to thread \texttt{tid} within the thread block, multiplying one non-zero element with one element from matrix $B$. All threads load all three operands for the FMA. This implementation resembles CSR-based methods, where non-zero elements within a row are accessible due to the CSR's row-based compressed storage format.


To reuse values of \texttt{B} in registers, Figure~\ref{fig:motivc1} maps three FMAs to one thread by unrolling the \texttt{i} iterator and reducing grid size. This is called thread coarsening across the unrolling factor (UF) \texttt{i}=4. This leads to coarser-grained threads, enabling the reuse of register accesses to matrix \texttt{B[k][j+tid]} as shown in lines~5--7. Scalarizing accesses to \texttt{B} is not shown for concise representation.
More coarsening is applied in Figure~\ref{fig:motivc2} by unrolling the \texttt{k} loop. This increases the reuse of elements from the second column of matrix $B$. Thread coarsening in Figures~\ref{fig:motivc1} and \ref{fig:motivc2} reduces the grid size to $M/4$, which can potentially under-utilize GPU cores depending on $M$.



  To show how coarsening improves each thread's performance by enabling more instruction-level parallelism, we use the Nsight GPU profiler to measure the utilization of the LSUs and FMUs. Figure~\ref{fig:motive_register} shows the profiling data for a sweep of \texttt{UFk}. Increasing the UFk  Profiling data is normalized over UFk=1. Increasing the UFk increase the register reuse on elements of Matrix A .As shown, both LSU utilization and FMA/cycles are improved by increasing the coarsening. This improvement also reduces the execution time of SpMM.
  As a result of coarsening, the performance of SpMM in Figure~\ref{fig:motivc2} has improved by 1.5 times over the code in Figure~\ref{fig:motivc0}.

 \begin{figure*}[!h]
\includegraphics[width=0.97\linewidth]{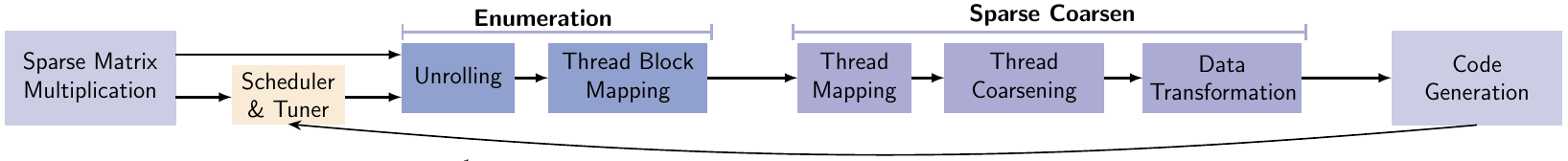}
     \caption{Overall view of enumerate-and-sparse-coarsen}
     \label{fig:overall}
 \end{figure*}

\section{Enumerate-and-Sparse-Coarsen}
The input to the enumerate-and-sparse-coarsen transformation is an SPMM code and the output of the transformation is an optimized \texttt{CUDA}/\texttt{C++} code. 

\subsection{Overview}

Figure~\ref{fig:overall} shows an overall view of the transformation. The transformation has two major parts: enumerate and sparse-coarsen. All steps of the transformation are applied to the dense SPMM code shown in Listing~\ref{lst:input}.
The enumerate part unrolls to expose sparsity pattern information by creating tiles with pre-determined access patterns and performing thread-block mapping. In the thread-block mapping, tiles and patterns are mapped to thread blocks to ensure load balance and minimize thread divergence.
Then, in the sparse-coarsen part, thread-mapping followed by thread coarsening is done to ensure register and L1 cache reuse and efficient warp-level parallelism. Finally, the code is transformed to work on a matrix storage format compatible with the coarsening schedule. The generated optimized SPMM code needs a transformed matrix \texttt{TA} as an input, as shown in line~\ref{lin:ta} in Listing~\ref{lst:output}. The executed code is also used combined with tuner to decide on efficient parameters for the scheduler (line~\ref{lin:sched} in Listing~\ref{lst:output}). 
This section discusses the details of each step of the transformation.

\begin{lstlisting}[caption={Input code}, label={lst:input},style=cpp]
for (i = 0; i < M; i++)
 for (k = 0; k < K; k++) 
  for (j = 0; j < N; j++) 
   if(A[i][k]) C[i][j] += A[i][k] * B[k][j];
\end{lstlisting}

\begin{lstlisting}[caption={Generated output code}, label={lst:output},style=cpp]
UFi,ThreadBlockSize,UFk,WarpTile=Scheduler("A100").get();|\label{lin:sched}|
TA = dataTransformer(A, UFi, UFk);|\label{lin:ta}|
int blockNo=TA.num_patterns * M / UFi;
spmmOpt<<<blockNo,ThreadBlockSize>>>(TA,B, C, UFi, UFj);
\end{lstlisting}

\subsection{Enumeration}

The enumeration phase of the transformation exposes the memory access patterns of a subregion of the matrix multiplication for load balancing and for potential compiler optimization. This phase involves two lowering steps: unrolling and thread-block mapping. These steps transform the input code in Listing~\ref{lst:input} into the code shown in Figure~\ref{fig:enum_example} using unrolling and thread block mapping steps.  

\begin{lstlisting}[caption={After unrolling and thread block mapping. Atomic instructions are not shown for a better illustration. i.e.  \texttt{UFi=4}}, label={lst:tiledunrolled},style=cpp]
int i = blockIdx.x * UFi; |\label{lin:threadmap}|
for(int j = 0 ; j < N ; j++)
 for(int k = 0 ; k < K ; k++){ |\label{lin:forresueC}|
  if(A[i+0][k]) C[i+0][j]+=A[i+0][k]*B[k][j]; |\label{lin:cond1}|   
  if(A[i+1][k]) C[i+1][j]+=A[i+1][k]*B[k][j]; |\label{lin:cond2}|     
  if(A[i+2][k]) C[i+2][j]+=A[i+2][k]*B[k][j]; |\label{lin:cond3}|   
  if(A[i+3][k]) C[i+3][j]+=A[i+3][k]*B[k][j]; |\label{lin:cond4}|
  }
\end{lstlisting}

\subsubsection{Unrolling}
To effectively leverage sparsity information and enable optimizations like register reuse, memory access patterns to arrays $A$, $B$, and $C$ must be known at compile time. Loop unrolling, a technique that converts memory accesses into scalar operations, is used for exposing accesses. Before applying unrolling, it is assumed all iteration of \texttt{i} in Listing~\ref{lst:input} is mapped to thread blocks using \texttt{map(i)}. Then \texttt{unroll("i", UFi)} directive is used to specify unrolling loop \texttt{i} with an unroll factor of \texttt{UFi}. Each row of $C$ is assigned to a thread block, as shown in line ~\ref{lin:threadmap} of Listing~\ref{lst:tiledunrolled}.
This unrolling exposes accesses to \texttt{A} through creating conditionals and also show common accesses to \texttt{B[k][j]}.
However, reusing $B$ is more challenging due to conditional statements dependent on the non-zero pattern of $A$, e.g., lines~\ref{lin:cond1}, \ref{lin:cond2}, \ref{lin:cond3}, and \ref{lin:cond4} in Listing~\ref{lst:tiledunrolled}. 

\begin{figure}[t]
    \raggedright
    \begin{subfigure}[b]{\columnwidth}
        \includegraphics[width=\linewidth]{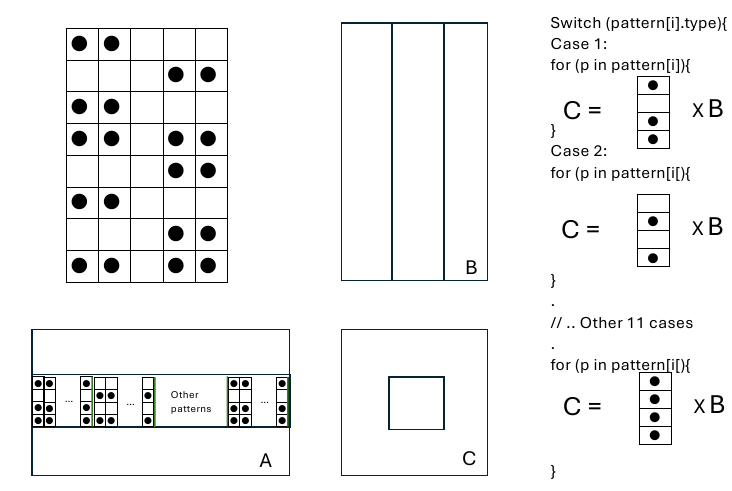}
        \label{fig:enum0}
    \end{subfigure}

    \begin{lstlisting}[label={lst:tiledunrolledfig4},style=cpp]
    int pattern = blockIdx.x % num_patterns;
    int row_panel = blockeIdx.x / num_patterns;
    switch(pattern){
    case 0:
     C[row_panel] += A[row_panel][pattern] * B;
     .
     .
    case 7:
     C[row_panel] += A[row_panel][pattern]  * B;
    }
}
    \end{lstlisting}

    \caption{Creating 2D tiles of $C$ using enumeration and tiling. Each row panel of $A$ has multiple patterns. The patterns in a dense matrix with similar pattern are shown adjacent for visualization. \texttt{num\_pattern} is 7 for UFi=3.}
    \label{fig:enum_example}
\end{figure}

To address the limited reuse of $B$ in listing~\ref{lst:tiledunrolled}, conditionals can be grouped together to enable simultaneous checking of multiple locations in $A$, thereby facilitating reuse of $B$. These unrolled statements enclosed by conditionals is called an \textit{enumerated block}. For example, lines~\ref{lin:condbeg}--\ref{lin:condend} in Listing~\ref{lst:enumerated} show an enumerated block where accesses to \texttt{B[k][j]} can be reused in a register.
While this approach improves register reuse by utilizing sparsity information, it limits the number of thread blocks to the number of row tiles in $A$, e.g. $m/4$ in Listing~\ref{lst:enumerated}. This can be problematic for short or small matrices. One potential solution is to create tiles for $B$ and map each 2D tile of $C$ to a thread block, as illustrated in Figure~\ref{fig:enum_example}. However, this approach may not fully utilize abundant GPU resources, as the $N$ in inference of neural network is not always large.
Additionally, threads within a thread block may diverge due to varying numbers of operations based on the non-zero pattern in $A$.

\begin{lstlisting}[caption={After conditionals are enumerated to build 15 enumerated blocks. i.e. \texttt{UFi=4}}, label={lst:enumerated},style=cpp]
int i = blockIdx.x * UFi; 
for(int j = 0 ; j < N; j++)
 for (int k = 0 ; j < K; k++){
  if(A[i][k] && A[i+1][k] && A[i+2][k] && A[i+3][k]){ |\label{lin:condbeg}|
   C[i+0][j] += A[i+0][k]*B[k][j];
   C[i+1][j] += A[i+1][k]*B[k][j];
   C[i+2][j] += A[i+2][k]*B[k][j];
   C[i+3][j] += A[i+3][k]*B[k][j];
  } |\label{lin:condend}|
  if(A[i][k] && A[i+1][k] && A[i+2][k] && !A[i+3][k]){
   C[i+0][j] += A[i+0][k]*B[k][j];
   C[i+1][j] += A[i+1][k]*B[k][j];
   C[i+2][j] += A[i+2][k]*B[k][j];
  }
  //...Other cases
 }
\end{lstlisting}

\subsubsection{Thread Block Mapping}

To ensure sufficient workloads for all thread blocks, the thread-block mapping step utilizes the created enumerated blocks to build workloads for thread blocks and mitigate thread divergence. The thread-mapping step enables the creation of sub-tiles in existing 1D tiles in $A$ by identifying columns with similar nnz locations within a tile of $A$. This approach resembles variable-sized 2D tiles of $A$, where each tile has a specific pattern. Figure~\ref{fig:enum_example}a shows an example of this tiling. As depicted, the 1D tile of $A$ with size $4$ is partitioned based on the non-zero patterns within the matrix.
For visualization purposes, patterns are shown adjacent to each other, but a list of pointers to the same pattern, \texttt{pattern}, is used internally. Corresponding to each pattern, there is an enumerated block as shown in Figure~\ref{fig:enum_example}b, which multiplies a predetermined pattern with matrix $B$.

Once 2D tiles of $A$ are created using sparsity information, threads are mapped to tiles with the same pattern to eliminate conditionals within thread blocks. 
 In element-wise sparse matrices, we observe a diverse range of patterns with nearly uniform frequency distribution. 
 Thus, we create thread blocks for all possible enumerated-block-tile combinations. 

\begin{lstlisting}[caption={After thread mapping in sparse-coarsen, i.e. \texttt{map(j,WarpTile*32)} where \texttt{WarpTile=1} and \texttt{UFi=4}}, label={lst:threadmap}, style=cpp]
int i = blockIdx.x * UFi; 
int tid = threadIdx.x % 32;
int jIndent = WarpTile * 32;
for(int j = 0 ; j < N;  j+=jIndent)
 for (int k = 0 ; k < K; k++){
   C[i+0][j+tid] += A[i+0][k]*B[k][j+tid];
   C[i+2][j+tid] += A[i+2][k]*B[k][j+tid];
   C[i+3][j+tid] += A[i+3][k]*B[k][j+tid];
 }
\end{lstlisting}

\subsection{Sparse Coarsen}

The sparse coarsen in the transformation, as shown in Figure~\ref{fig:overall}, has three steps. Sparse coarsen starts with the enumerated code shown in Figure~\ref{fig:enum_example}b and applies thread mapping, coarsening, and data transformation to generate the optimized CUDA code for SPMM. This section illustrates the lowering steps of sparse coarsen and how they are applied to an enumerated block.

\subsubsection{Thread mapping} 
The thread mapping step maps consecutive loop iterations to a warp, which consists of 32 threads. This is achieved using the \texttt{map(j,W)} directive, where \texttt{W} specifies the number of consecutive iterations of $j$ to be mapped to a warp. The lowering process applies this by replacing the loop increment with \texttt{32} and replacing the loop variable \texttt{j} with \texttt{j+tid} in the loop body. Variable \texttt{tid} is the thread ID within a warp. An example of thread mapping applied to the enumerated block in Figure~\ref{fig:enum_example}b is shown in Listing~\ref{lst:threadmap}. The choice of loop iterator for mapping affects the number of atomic operations. For instance, \texttt{map(k,32)} in Figure~\ref{fig:motivc0} requires atomic writes to $C$ because multiple threads write to the same location. However, this can improve load balance when the matrix is highly sparse and more workload is needed. 





\begin{lstlisting}[caption={After thread coarsening in sparse-coarsen. i.e.
\texttt{map(j,WarpTile)} where \texttt{WarpTile=2} and 
\texttt{map(k,UFk)} where \texttt{UFk=2} and UFi=4}, label={lst:thread_coarsen}, style=cpp]
int i = blockIdx.x * UFi; 
int tid = threadIdx.x % 32;
int jIndent = WarpTile*32;
int kIndent = UFk;
for(int j = tid ; j < N; j+=jIndent){
  float c00,c01;
  float c20,c21;
  float c30,c31;
  for (int k = 0 ; k < K; k+=kIndent){
    c00 += A[i+0][k]*B[k][j];
    c20 += A[i+2][k]*B[k][j];
    c30 += A[i+3][k]*B[k][j];
    c01 += A[i+0][k]*B[k][j+32];
    c21 += A[i+2][k]*B[k][j+32];
    c31 += A[i+3][k]*B[k][j+32];
    c00 += A[i+0][k+1]*B[k+1][j];
    c20 += A[i+2][k+1]*B[k+1][j];
    c30 += A[i+3][k+1]*B[k+1][j];
    c10 += A[i+0][k+1]*B[k+1][j+32];
    c21 += A[i+2][k+1]*B[k+1][j+32];
    c31 += A[i+3][k+1]*B[k+1][j+32];
}
 AtomicAdd(C[i+0][j],c00);AtomicAdd(C[i+0][j+32],c01);|\label{lin:atomicb}|
 AtomicAdd(C[i+2][j],c20);AtomicAdd(C[i+2][j+32],c21);
 AtomicAdd(C[i+3][j],C30);AtomicAdd(C[i+3][j+32],c31);|\label{lin:atomice}|
}}

\end{lstlisting}

\subsubsection{Thread Coarsening} This step finds a trade-off between warp-level and instruction-level parallelism. It maps a group of operations to each thread in a warp to maximize data reuse through register and L1 cache reuse. This lowering step has two stages:
I) coarsening threads by unrolling a loop iterators so each thread performs more operations. 
II) adding synchronization to prevent race conditions. Two types of write-after-write (WAW) interactions can occur: within a warp and across thread blocks. Warp-level WAWs are handled using warp-level reduction, while atomic add operations, e.g., \texttt{AtomicAdd} are used to ensure correctness across thread blocks. 
Listing~\ref{lst:thread_coarsen} shows the output of thread coarsening stage applied to Listing~\ref{lst:threadmap}. As shown, \texttt{unroll("k", 2)} followed by atomic add insertion are done in lines~\ref{lin:atomicb}--\ref{lin:atomice} in Listing~\ref{lst:thread_coarsen}.

\begin{lstlisting}[caption={After data transformation.
i.e.
\texttt{map(j,WarpTile)} where \texttt{WarpTile=2} and 
\texttt{map(k,UFk)} where \texttt{UFk=2} and UFi=4. ANNZ is the compressed array that stores the non-zero elements of the transformed matrix A.
NPP is the row panel pointer array, indicating the start of each row panel in ANNZ. RPP contains the column indices corresponding to the enumerated columns of matrix A for each row panel.}, label={lst:data_transform}, style=cpp]
int i = blockIdx.x * UFi; 
int tid = threadIdx.x % 32;
int jIndent = WarpTile*32;
int kIndent = UFk;
for(int j = th ; j < N; j+=j_indent){
  float c00,c01;
  float c20,c21;
  float c30,c31;
  int t_nnz = NPP[i]
  for (int k = RPP[i] ; k < RPP[i+1]; k+=k_indent){ |\label{lin:loop_col}|
    int br0=Cols[k]; int br1=Cols[k+1]; |\label{lin:offset}|
    c00 += ANNZ[t_nnz+0]*B[br0][j]; |\label{lin:pnnzb}|
    c20 += ANNZ[t_nnz+1]*B[br0][j];
    c30 += ANNZ[t_nnz+2]*B[br0][j];
    c01 += ANNZ[t_nnz+3]*B[br0][j+32];
    c21 += ANNZ[t_nnz+4]*B[br0][j+32];
    c31 += ANNZ[t_nnz+5]*B[br0][j+32];
    c00 += ANNZ[t_nnz+0]*B[br1][j];
    c20 += ANNZ[t_nnz+1]*B[br1][j];
    c30 += ANNZ[t_nnz+2]*B[br1][j];
    c10 += ANNZ[t_nnz+3]*B[br1][j+32];
    c21 += ANNZ[t_nnz+4]*B[br1][j+32];
    c31 += ANNZ[t_nnz+5]*B[br1][j+32]; |\label{lin:pnnze}|
    t_nnz+=6;
 }
  AtomicAdd(C[i+0][j],c00);AtomicAdd(C[i+0][j+32],c01);
  AtomicAdd(C[i+2][j],c20);AtomicAdd(C[i+2][j+32],c21);
  AtomicAdd(C[i+3][j],C30);AtomicAdd(C[i+3][j+32],c31);
 }
\end{lstlisting}

\subsubsection{Data transformation} 
After determining the schedule, including parameters from both the enumeration and coarsening phases, the non-zero elements of $A$ are stored in a compressed format to remove zero elements.
Using compressed structures to reduce the size of weight matrices is crucial. The data transformation step, the final step of the lowering process, removes zero elements and stores non-zero elements based on the computation schedule. The primary performance objective of this re-storage is to improve coalesced memory access patterns in SpMM. Lines~\ref{lin:pnnzb}--~\ref{lin:pnnze} in Listing~\ref{lst:data_transform} illustrates how the SpMM code is transformed to operate on a compressed array \texttt{ANNZ}. After data transformation, all accesses to matrices $A$, $B$ and $C$ are coalesced.  


%


    


\subsection{ Scheduler \& Tuner}
The goal of the scheduler is to determine the scheduling parameters used in the lowering steps of enumerate-and-sparse-coarsen. This section discuss the tuning approach that is used to builds the schedule. 
Each schedule in the enumerate-and-sparse-coarsen is represented with list of directives, \texttt{Unroll, Map, Map, Unroll, Transform}, one for each step of the transformation. Considering different unrolling factors and mapping to the three loops, the search space can grow large for tuning. However, we use a profiling-based tuning approach to reduce the tuning space. As a result, we use parameters, \texttt{UFi, ThreadBlockSize} as unrolling factor and thread block size in the enumerate and \texttt{WarpTile, UFk} as the warp tile and coarsening factor of the sparse coarsen. We assume data transformation is always needed.

\begin{figure}[h]
    \centering
    \includegraphics[width=\linewidth]{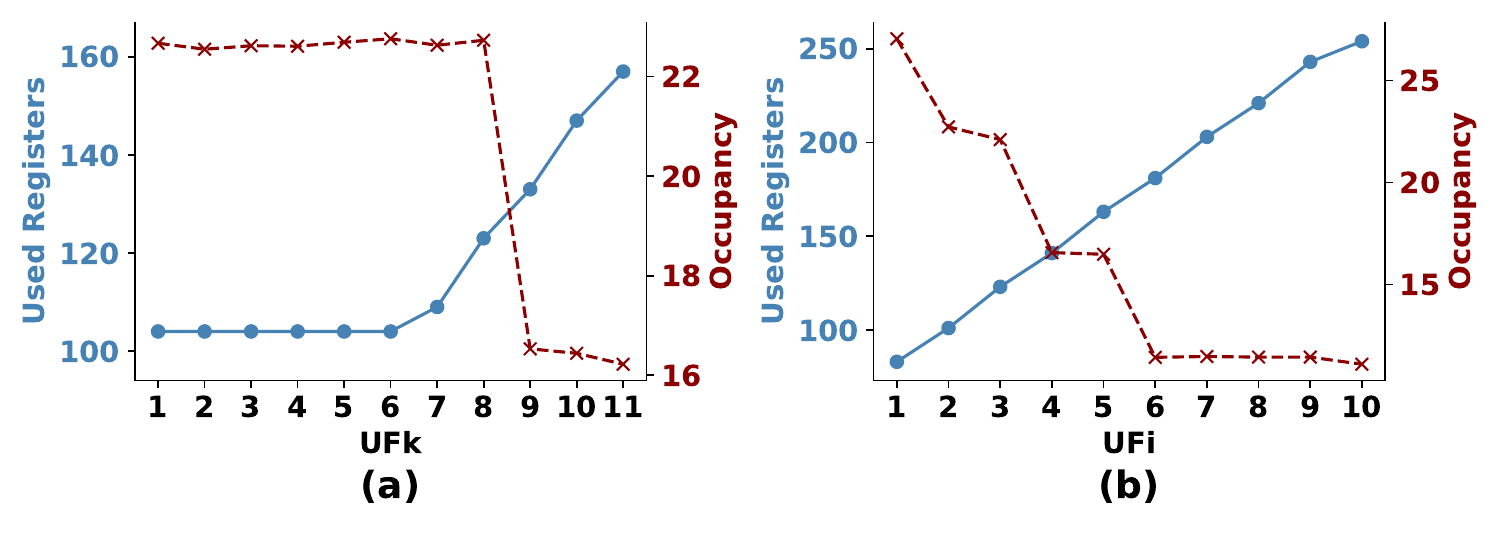}
    \caption{ Profiling different values of UFk and UFi to show the trade-off between register reuse and occupancy. The profiling is used to reduce the tuning space for the scheduler.
    }
    \label{fig:profiling:sweep}
\end{figure}

We use architecture information to find a reasonable and small schedule space for tuning. Figure~\ref{fig:profiling:sweep} shows the tuning range of \texttt{UFi} and \texttt{UFk} are selected. This figure is generated by running a sweep of values for all unique dimension and the 5 sparsity ranges in DLMC, which makes 60 matrices. We observe a similar trend to Figure~\ref{fig:profiling:sweep} across all matrices. Figure~\ref{fig:profiling:sweep} is illustrated for a matrix with sparsity ratio of 70\% and dimension 2048$\times$1024. As shown, only \texttt{UFi} and \texttt{UFk} values are selected that keep the occupancy above 16 and 12, respectively. To select the size of \texttt{WarpTile} and \texttt{ThreadBlockSize}, the tuning parameters are selected to be a multiple of 32 and smaller than $N$.


\subsection{Code Generation} 
\label{sec:codegen}
We implemented the Enumerate-and-Sparse-Coarsen transformation as a Python-based source-to-source code generator that produces \texttt{C++/CUDA} code. As shown in Listing~\ref{lst:output}, the generated code comprises two parts:
I) the \texttt{dataTransformer} function in \texttt{C++} that converts the input sparse matrix $A$ to the compressed form according to the selected schedule. II) the optimized SpMM code in \texttt{CUDA} that performs the multiplication on the compressed matrix $A$. 










\subsubsection*{Data transformer }  The input to the data transformation stage is the selected schedule, and the output is a \texttt{dataTransformer(A)} function. This function takes the dense matrix $A$ as input and converts it into a new format based that the lowerd code can iterate over.   
The data transformer function finds enumerated tile based on \texttt{UFi} and \texttt{UFk} and put them in consecutive locations in the compressed format. 
The transformation iterates over dense matrix $A$ once, making the computational complexity of the transformation as $O(M*K)$.
The data transformer and the SpMM code are generated separately to enable the reuse of the compressed matrix multiple times. For neural network inference, the non-zero locations and elements remain constant throughout the inference process, allowing for the reuse of the compressed format. 


\subsubsection*{SpMM optimized code} 
We represent the iteration space and memory access patterns in matrix multiplication code using sets and access functions. The matrix multiplication is represented with iterations space of $\mathcal{I}$ and three access functions $f$, $g$, and $h$ for accesses to $C$, $A$, and $B$, respectively. For example, the input code in Listing~\ref{lst:input} is shown with integer tuples $\mathcal{I}={0\leq i<M, 0\leq j<N, 0\leq k<K}$. Access functions are $f=i*M+j, g=i*M+k, h=j*K+k$. 
Since all transformations prior to data transformation are applied to a dense storage format, access functions are defined as affine or linear combinations of loop variables. For each function, a dictionary of loop iterators is created, where the integer value associated with each iterator represents its coefficient in the generated code. After data transformation, access functions may no longer be affine. However, the non-affine part can be separated as an offset, allowing us to represent non-affine functions as a combination of a variable and an affine access function for final code generation. All statements, including switch statements and multiply-add operations, within loop bodies are stored as a list of statements in lexicographical order, labeled with scope numbers. 

The two important lowering stages in enumerate-and-sparse-coarsen are \texttt{unroll} and \texttt{map}. 
\texttt{unroll(i,uf)}: 
The unrolling directives modify the loop increment to the current unrolling factor $uf$ and then unroll statements involving $i$ by replacing $i$ with $i+0$ up to $i+uf-1$. 
%
\texttt{map(i,s)}: 
The mapping directive modifies the loop increment of $i$ to $s$ if $s$ is smaller than the loop bound. For instance, in the thread mapping phase of the sparse-coarsen, the loop increment is smaller than the warp size of 32. Consequently, the selected loop iterator $j$ in is replaced with \texttt{j+tid}, where \texttt{tid} is the thread ID. However, in thread-block mapping, $s$ represents the grid size, which is typically a large value exceeding the number of rows and enumerated blocks. In this case, the mapping removes the loop iterator \texttt{i} and replaces it with \texttt{blockIdx.x}. 

While the code size for feasible schedules does not become a significant issue, the compilation time of the generated code using the CUDA compiler can be substantial. We propose a code compaction approach based on non-zero elements to reduce code size. Instead of generating a function for each enumerated block's body, which would result in $2^m$ functions for a tile size of $m$, we generate only $m$ code snippets. In other words, the number of non-zero accesses within an enumerated block is used as an identifier to generate a function. To achieve this, we pass the offset to access functions $g$ as input to the function. 

\noindent
\textit{\textbf{Safety}} 
Since matrix multiplication operations are inherently independent, the only potential dependence is a write-after-write dependency, which can lead to race conditions. The enumerate-and-sparse-coarsen transformation mitigates this issue by using either atomic operations or reduction for all writes to the output matrix or shared variables, ensuring the correctness of the transformation.

\section{Experimental Results}
In This section the performance of the enumerate-and-sparse-coarsen method is evaluated for SPMM. Overall enumerate-and-sparse-coarsen outperforms cuSparse and CuBlas baselines across a range of bCols on DLMC \cite{gale2020sparse} and sparse attention matrices.  


\begin{figure*}[!t]
    \centering
    \includegraphics[width=\linewidth]{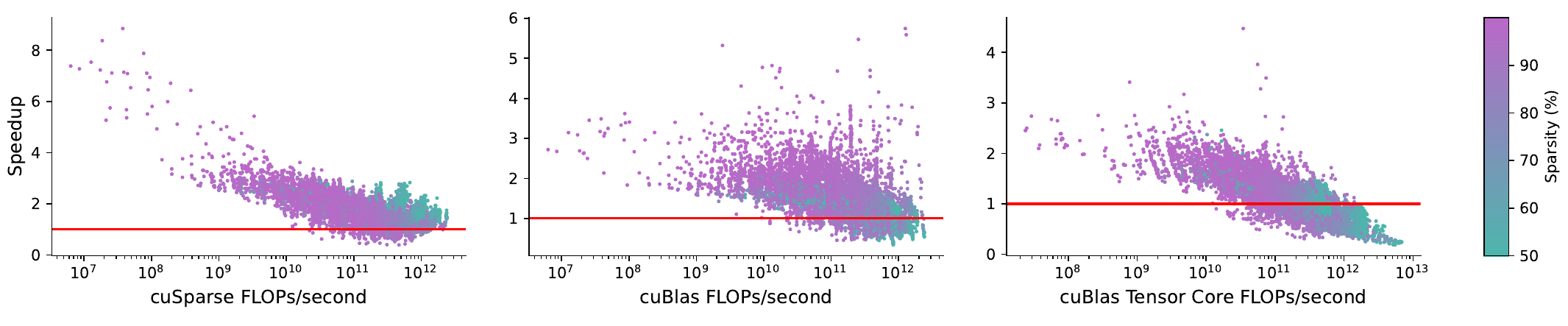}
    \caption{The performance of enumerate-and-sparse-coarsen for different bCols for three different baselines. }
    \label{fig:efffect:of:bcol}
\end{figure*}

\subsection{Setup}
\subsubsection{Dataset} To evaluate the performance of the proposed enumerate-and-sparse-coarsen transformation, we conducted experiments using unstructured matrices from the DLMC, including weight matrices derived from ResNet50 and Transformer models. These matrices present diverse computational challenges, enabling a comprehensive evaluation of the proposed techniques.


For unstructured matrices, we evaluated 3,608 matrices from DLMC~\cite{gale2020sparse}, all with sparsity greater than or equal to 50\%. These matrices vary in size, with the smallest being 64$\times$64 and the largest 33,288$\times$512. The most common sizes in the dataset are 512$\times$512, 2048$\times$512, and 512$\times$2048. This dataset includes matrices extracted from pruned Transformer architectures and ResNet50 models.

In the case of ResNet50, the im2col operation is applied to unfold convolutional filters, transforming convolutional layers into matrix multiplication problems. The pruning of weight matrices was performed using several unstructured techniques, including variational dropout \cite{kingma2015variational}, $l_0$ regularization \cite{louizos2017learning}, random pruning, and an extended version of magnitude-based pruning \cite{zhu2017prune}. The dataset comprises a total of 59 models.

\subsubsection{Target Platform}  
The experiments were conducted on an NVIDIA A100 GPU using CUDA version 12.2, ensuring compatibility with the underlying architecture and modern GPU features. All implementations were developed in \texttt{C++11} and the code generator is developed in Python.
The A100 GPU, a data-center-grade accelerator, features 108 Streaming Multiprocessors (SMs) operating at a base clock of 765 MHz. 
Each SM provides up to 192 KB of configurable L1/shared memory, while the 40 MB L2 cache supports efficient data reuse—crucial for memory-intensive operations.

\subsubsection{Methods} 
Each experiment is repeated twenty times, and the median runtime is reported to ensure consistency. Initial warm-up iterations are also included to stabilize the cache state. The experiments are conducted multiple times to ensure reproducibility of performance measurements. We compare the enumerate-and-sparse-coarsen transformation with cuBLAS dense MatMul with and without tensor cores denoted as cuBlas and cuBlas Tensor Core, respectively. We also use cuSparse CSR as our sparse baseline. Throughout the section we use $bCol$ for columns of $B$ or $N$.  
\subsection{Performance Evaluation}
\label{sec:perf}


\begin{table}[!t]
\centering
\begin{tabular}{|c|ccc|}
\hline
\textbf{} & 32 & 64 & 128 \\
\hline
\multicolumn{4}{|c|}{\textbf{Speed-up}} \\
\hline
Vs cuBLAS   & 1.40 & 1.48 & 1.54 \\
Vs cuBLASTC & 1.17 & 1.09 & 1.06 \\
Vs cuSparse & 1.85 & 1.78 & 1.60 \\
\hline
\multicolumn{4}{|c|}{\textbf{Faster Matrices (\%)}} \\
\hline
Vs cuBLAS   & 85.54 & 87.38 & 89.67 \\
Vs cuBLASTC & 73.77 & 64.88 & 59.62 \\
Vs cuSparse & 96.05 & 95.57 & 95.11 \\
\hline
\end{tabular}
\caption{Geometric mean of speedups and percentage of faster execution for DLMC matrices for A100 GPU.}
\label{tab:spmm:speedup:groupedbcols:faster}
\end{table}


This section discusses the performance of the enumerate-and-sparse-coarsen on SPMM across datasets. Data Type that is used in all experiments is float32.

Figure~\ref{fig:efffect:of:bcol} illustrates the overall performance of enumerate-and-sparse-coarsen for SpMM on DLMC matrices across both small and large bCols. Enumerate-and-sparse-coarsen achieves a geometric mean speedup of 1.76 and 2.28 over cuSparse and cuBlas, respectively, for these DLMC matrices. As the figure highlights, the transformation yields greater speedups with smaller bCols compared to larger ones.

Table~\ref{tab:spmm:speedup:groupedbcols:faster} compares the performance of the enumerate-and-sparse-coarsen transformation across varying bCols against cuBLAS and cuSparse baselines using the DLMC matrices dataset. The ``Speed-up'' values represent the geometric mean of speedup factors observed for the small and large bCol sizes presented in the table. Consistent with Figure~\ref{fig:efffect:of:bcol}, the data in Table~\ref{tab:spmm:speedup:groupedbcols:faster} underscores the dominance of speedup figures in smaller bCols. The most significant geometric mean speedup is achieved over cuSparse, while the lowest is seen with cuBlas tensor cores for large bCols, where the efficiency of tensor cores becomes apparent.

The ``Faster Matrices'' section of Table~\ref{tab:spmm:speedup:groupedbcols:faster} reveals the percentage of matrices where enumerate-and-sparse-coarsen demonstrates superior performance compared to the baselines. Across both bCols and baselines, the transformation outperforms all baselines in a significant majority of cases, ranging from 60\% to 100\% of the matrices. When compared to cuBlas with large bCols, an interesting trend emerges. For the tensor core variant, increasing bCols enhances cuBlas performance, likely due to the effectiveness of the tensor cores. Conversely, for the non-tensor core variant, the trend is reversed, likely stemming from the increased redundant computations on zero values as bCol size grows.

\subsection{Ablation Study}

\subsubsection{Performance breakdown}
\begin{figure*}[!t]
    \centering
    \includegraphics[width=0.8\linewidth]{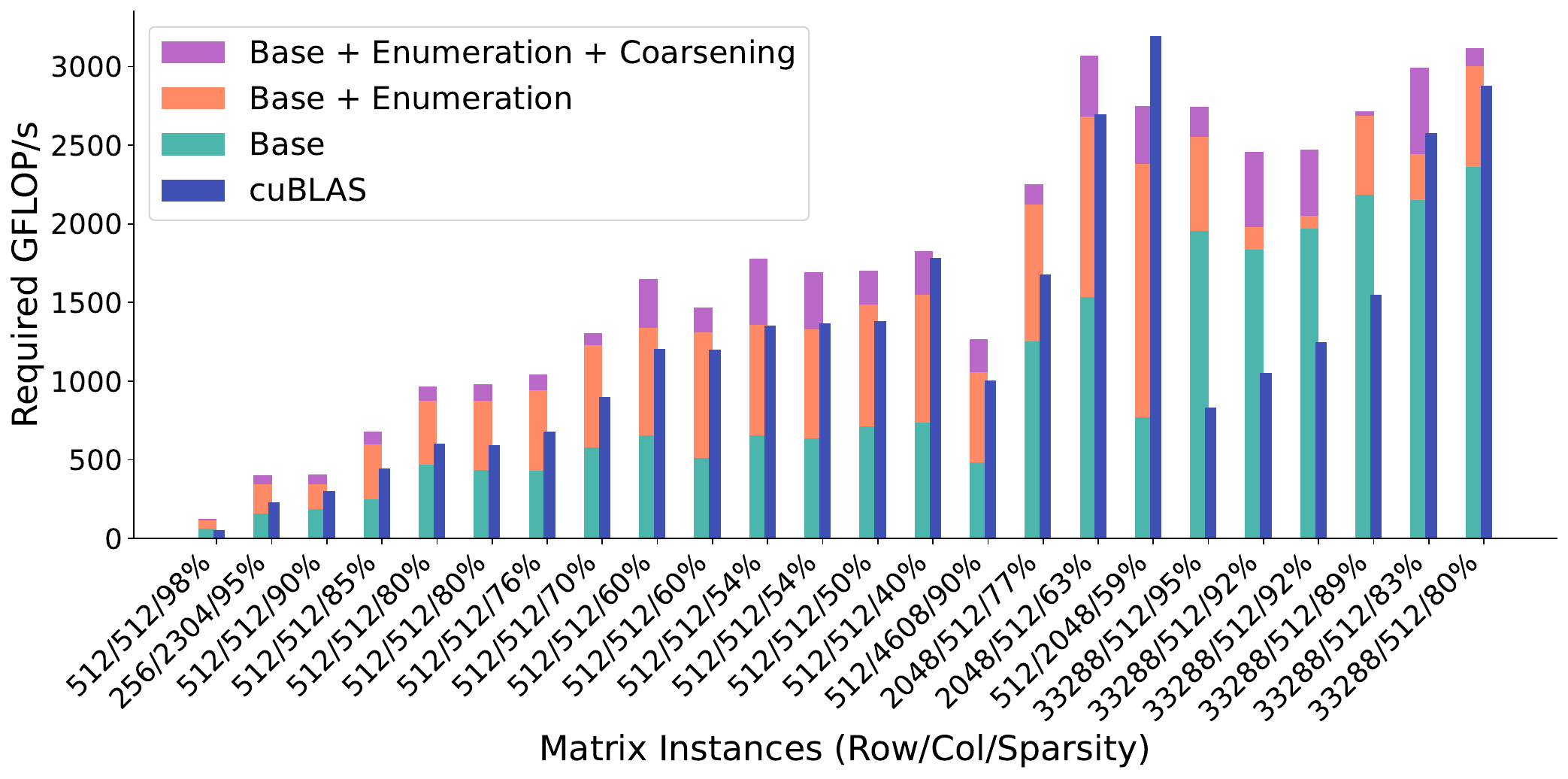}
    \caption{Performance breakdown of enumerate-and-sparse-coarsen}
    \label{fig:lowering}
\end{figure*}

Figure~\ref{fig:lowering} illustrates the performance breakdown of the different optimization steps in enumerate-and-sparse-coarsen, compared to cuBLAS, across 25 random matrices. The x-axis displays the matrix properties, and the y-axis indicates the required GFLOP/s.

The Base configuration corresponds to the initial implementation (as depicted in Figure~\ref{fig:motivc0}), where no coarsening is applied and the dense matrix format is used. The Base + Enumeration step enhances the baseline by enumerating over sparsity patterns. The Base + Enumeration + Coarsening step incorporates all previous optimizations along with coarsening.
It is important to note that both enumeration and coarsening significantly contribute to performance improvement. As shown in the figure, for matrices with similar sizes and sparsity levels, the effect of each optimization step is generally consistent. In some cases, enumeration alone is insufficient to outperform cuBLAS—demonstrating that coarsening is essential for achieving higher performance.

\subsubsection{Cost model}

This section discusses a few examples of optimal schedules used in Enumerate-and-Sparse-Coarsen and how they consider different properties of sparse MatMul.
Each schedule includes the unrolling factor of i (UFi), the unrolling factor of k (UFk), the number of columns of B assigned to each thread (WarpTile), and the thread block size (threadBlockSize). The unrolling factors are more related to the dimensions of A, while the last two are more related to the bCol size. Table~\ref{tab:config} shows the best configuration and the speedup achieved with that schedule for all unstructured DLMC matrices. As can be seen, if only one schedule were to be chosen, we would lose almost 10\% in performance compared to the results presented in Table~\ref{tab:spmm:speedup:groupedbcols:faster}.


\begin{table}[!t]
\centering
\begin{tabular}{|c|c|c|c|}
\hline
 bCol & Schedule & cuSPARSE & cuBLAS \\ \hline
32 & 4-7-1-32 & 1.73 & 1.32 \\ \hline
64 & 3-7-2-32 & 1.69 & 1.38 \\ \hline
128 & 3-8-2-64 & 1.62 & 1.55 \\ \hline
\end{tabular}
\caption{Performance comparison between the best schedule, cuBLAS, and cuSPARSE for various bCol values on NVIDIA A100 GPU. The Schedule is UFi-UFk-WarpTile-ThreadBlockSize
}
\label{tab:config}
\end{table}

\begin{figure}[!t]
    \centering   \includegraphics[width=\linewidth]{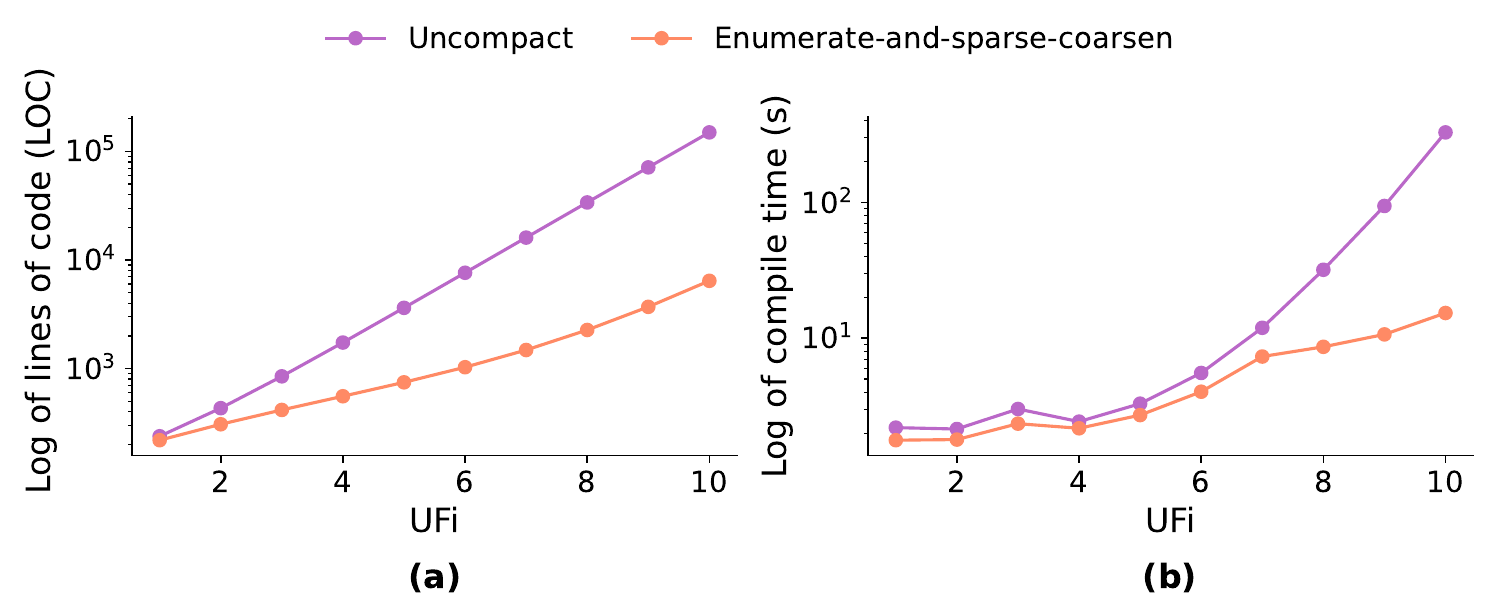}
    \caption{(a) lines of code and (b) compile time for the enumerate-and-sparse-coarsen version compared to the uncompressed code size}
    \label{fig:compile:time}
\end{figure}
\begin{figure}[!t]
    \centering
    \includegraphics[width=\linewidth]{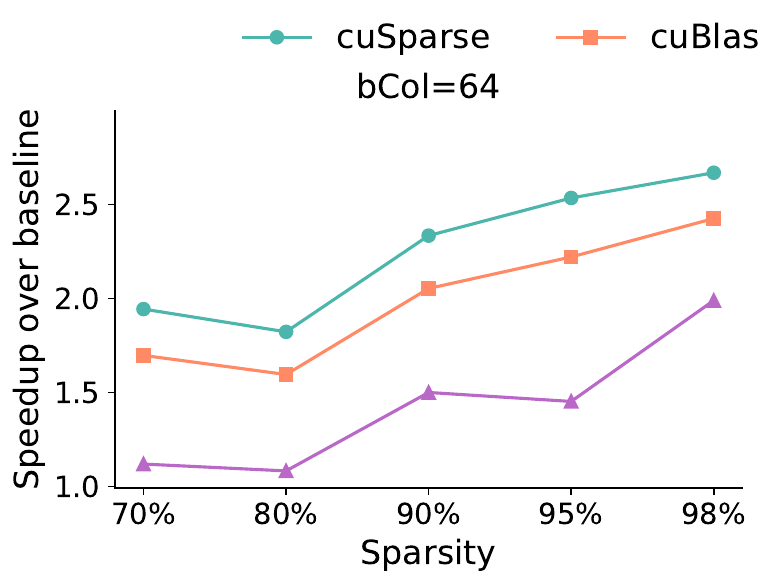}
    \caption{Sweep showing speedup across varying sparsity levels for a 512×512 square matrix with two bCol sizes: 64 (large) and 4 (small).}
    \label{fig:matrix:sweep}
\end{figure}

\begin{figure}[!t]
    \centering
    \includegraphics[width=\linewidth]{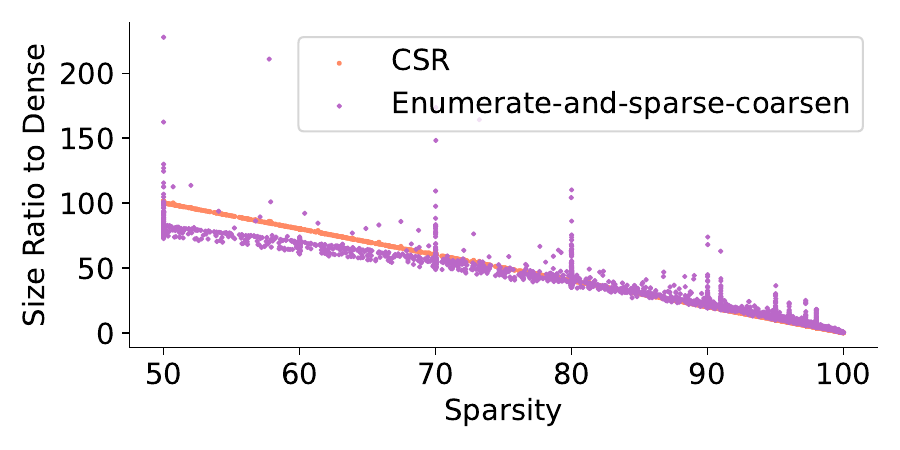}
    \caption{Bytes required for CSR and enumerate-and-sparse-coarsen-format, normalized by dense (Lower is better).}
    \label{fig:matrix:compressed}
\end{figure}




\subsubsection{Compile-Time Analysis}

This section explores how the compaction strategy detailed in Section~\ref{sec:codegen} reduces compile-time. Figure~\ref{fig:compile:time}-b presents the \texttt{nvcc} compile-time for various schedules, both with and without code compaction. The time measurements were taken using the bash timer. Notably, the compile-time associated with enumerate-and-sparse-coarsen can be entirely eliminated since tuning is performed only once per architecture and for each value of bCol. Given that tuning is a one-time process per architecture, the remaining compile-time is solely the duration spent in the CUDA compiler, specifically \texttt{nvcc}.

\subsubsection{Sparsity Sweep} 
For a concise view of Enumerate-and-Sparse-Coarsen performance, refer to Figure~\ref{fig:matrix:sweep}, which illustrates a sparsity sweep for the selected bCol of 64. As the sparsity ratio increases, the transformation's performance improves across all baselines. Furthermore, consistent with the findings in Section~\ref{sec:perf}, the generated code consistently outperforms all other baselines.





\subsection{Data structure size}

Figure~\ref{fig:matrix:compressed} shows the storage requirements of enumerate-and-sparse-coarsen compared to the CSR storage format, both normalized against a dense matrix format. Sparse Unroll and Coarsen outperforms CSR in storage efficiency for matrices with sparsity levels ranging from around 50\% to 80\%. This is because it reduces indirection to one per merged enumerated block, whereas CSR maintains an indirection per nonzero element. However, as sparsity approaches 99\%, the method’s storage efficiency diminishes due to the overhead from metadata, making CSR a more favorable option.

\subsection{effect of pruning method and models}
This section analyzed the SpMM performance of generated code for matrices pruned from transformer and RN50 models within the DLMC dataset. Overall, the transformer model demonstrated superior and more consistent performance compared to RN50.
Within the transformer models, the transformation yielded the most significant performance gains, particularly with L0 regularization. Furthermore, for transformer matrices, random and magnitude pruning methods achieved better average FLOPs/seconds compared to L0 regularization and variational dropout. The greater variability observed in RN50 performance, compared to transformers, can be attributed to the considerably wider range of matrix dimensions present in RN50, encompassing 21 different sizes versus 4 sizes in transformers. 

\section{Related Work}

\subsubsection*{Sparse neural network inference}
Several pruning algorithms, such as those proposed in \cite{kingma2015variational,louizos2017learning,zhu2017prune,gray2017gpu}, have been developed to reduce the size of machine learning models. Pruning algorithms can be categorized based on their granularity: element-wise, vector-wise, or block-wise~\cite{gray2017gpu}. While element-wise pruning, as described in \cite{kingma2015variational,louizos2017learning,zhu2017prune}, often yields higher accuracy, it can be less efficient to execute on hardware. One of the major challenges is to turn the reduced number of FLOPs into performance in sparse neural network, where often dense matrix multiplication such as cuBLAS~\cite{naumov2010cusparse} are used. This paper focuses on a code generation framework for the SPMM operation, specifically targeting element-wise pruning approaches.

\subsubsection*{Libraries}
 Several domain-specific libraries have been developed to accelerate SPMM~\cite{fu2024jitspmm} and SPMV~\cite{you2022vectorizing} on CPUs. These methods rely on efficient tiling and load balancing strategies to fully utilize all CPU cores.
The GPU programming model, which allows for finer-grained task creation, requires different strategies for optimizing SPMM and SpMV. Several libraries have been designed for GPUs to accelerate both SPMV~\cite{merrill2016merge,naumov2010cusparse,greathouse2014efficient,yang2018design} and SPMM~\cite{xiaflash,hong2018efficient,naumov2010cusparse,hong2019adaptive,gale2019state,yang2018design}. Libraries like AsPT \cite{hong2019adaptive} and cuSparse \cite{naumov2010cusparse} are optimized for high-sparsity scenarios,e.g. 90\% and higher. However, they can be less efficient than dense matrix multiplication libraries like cuBLAS \cite{naumov2010cusparse}, especially for denser matrices, due to the regular access patterns and tensor core utilization in cuBLAS.
There are also libraries, such as EC-SpMM \cite{lin2023ec}, Sputnik \cite{gale2019state}, and FlashLLM \cite{xiaflash}, are specifically designed for DNNs. These methods employ 1D tiling~\cite{gale2019state} or 2D tiling~\cite{xiaflash,lin2023ec} to ensure load balance across all GPU SMs. They also restore the sparse matrix to improve coalesced memory access.


\subsubsection*{Compilers}
Compilers like Halide \cite{ragan2013halide}, Exo \cite{ikarashi2022exocompilation}, TVM \cite{chen2018tvm}, and Polyhedral compilers \cite{vasilache2018tensor} build schedule plans from user hints and apply them. Our approach also decouples code transformation from scheduling, but it uses a tuning-based scheduler to determine the optimal schedule. However, these compilers primarily focus on dense computations.
%
Several compilers\cite{wilkinson2023register,strout2018sparse,fu2024jitspmm,kjolstad2017tensor,won2023waco,ahrens2024finch,ghorbani2024compressing,dias2022sparselnr,horro2022custom,cheshmi2022vectorizing,cheshmi2017sympiler,cheshmi2022transforming} provide efficient parallel SPMM and SPMV implementations. TACO~\cite{kjolstad2017tensor}, SparseLNR~\cite{dias2022sparselnr}, and COMET~\cite{COMET:LCPC-20} provides a tensor expression abstraction and WACO~\cite{won2023waco} builds an auto-scheduler for optimized execution of tensor expression on CPUs. Sparse polyhedral framework (SPF)~\cite{strout2018sparse,mohammadi2019extending} extends polyhedral frameworks to non-affine accesses in sparse codes.  
FINCH~\cite{ahrens2024finch} and StructTensor~\cite{ghorbani2024compressing} are programming languages that allow describing patterns as part of grammar and thus enabling compile-time optimization. 
Unroll-and-Sparse-Jam \cite{wilkinson2023register} exposes access patterns of sparse codes through an enumeration process, similar to Deep Jam \cite{carribault2005deep}. 
Unroll-and-Sparse-Jam applies this technique to SPMM by unrolling conditionals to create multiple enumerated blocks, which are then vectorized using AVX intrinsics. It also uses zero-padding to reduce code size.
Enumerate-and-Sparse-Coarsen also benefits from enumerated blocks, but faces the challenge of avoiding thread divergence as the number of conditionals increases after enumeration. Our approach addresses this by using a novel thread block mapping strategy that ensures thread blocks do not contain conditionals and distributes blocks uniformly across thread blocks. 

%



Fractal \cite{guan2024fractal}, SparseTIR \cite{ye2023sparsetir}, and SparTA \cite{zheng2022sparta} are GPU-focused compilers that generate efficient code for sparse neural network. Fractal, in particular, leverages block structures to trade off accuracy and performance. SparseTIR provides a composable abstraction for converting tensor operations into loops and applying transformations. SparTA considers the non-zero patterns of weight matrices during code generation, leading to efficient code for block-structured scenarios. However, for element-wise pruning, where block structures may not exist, SparTA relies on 1D tiling.
While these GPU compilers prioritize efficient GPU resource utilization through increased workload and coalesced access, register reuse is often limited to block-structured scenarios. Enumerate-and-sparse-coarsen addresses this limitation by considering both tile-level and register-level optimizations for element-wise pruning methods.

\subsubsection*{Coarsening in GPU}
Thread coarsening \cite{barua2018cost,magni2014automatic,liu2022indigo,hong2018efficient} is a well-known technique for improving GPU code performance by reducing occupancy in exchange for increased register reuse.
Threads coarsening can be applied at the thread-block-level and thread-level \cite{stawinoga2018predictable} to improve load balance and locality, respectively.
Thread coarsening has been applied to SPMM and SpMV in GPU libraries such as \cite{hong2018efficient}. However, its primary purpose has been to reduce the overhead of shuffle instructions and improve coalesced access to $B$. Thread coarsening is also used to enable loop fusion in Tile Fusion~\cite{salehi25loop,dezfuli2024improving} to enhance cache reuse. Leveraging thread coarsening to enhance register reuse, as explored in sparse coarsening, is a novel aspect not extensively investigated in previous coarsening approaches.

\section{Summary and Conclusion}
This paper introduces enumerate-and-sparse-coarsen, a novel transformation for SPMM. This transformation emphasizes thread-level optimization, facilitating register reuse. The transformation has two steps. The enumeration step of the transformation exposes memory accesses at compile-time and also ensures load balance for thread blocks. The sparse-coarsen step enables register reuse by mapping multiple FMA operations to one thread. The transformed SpMM operation achieves gmean speedups of 1.76$\times$ and 2.28$\times$ over cuBLAS and cuSparse, respectively. 
\begin{acks}
This work is supported by NSERC discovery grants (RGPIN-2023-04897, DGECR-2023-00133), NSERC alliance grant (ALLRP 586319-23), and the Digital Research Alliance of
Canada (www.alliancecan.ca).
\end{acks}

\bibliographystyle{ACM-Reference-Format}
\bibliography{sample-base}



\end{document}